\documentclass[twocolumn,aps]{revtex4-1}
\usepackage[latin9]{inputenc}
\usepackage{amsmath}
\usepackage{amssymb}
\usepackage{graphicx}
\usepackage{esint}
\usepackage{color}
\usepackage{import}

\makeatletter
\usepackage{epstopdf}

\usepackage{verbatim}

\newcommand{\ev}[1]{\left\langle#1\right\rangle}
\newcommand{\tr}[1]{\operatorname{\text{Tr}}\left\{#1\right\}}
\newcommand{\tra}[2]{\operatorname{\text{Tr}_{#1}}\left\{#2\right\}}
\newcommand{\de}[2]{\frac{\mathrm d^{#1}}{\mathrm d{#2}^{#1}}} 
\newcommand{\prt}[2]{\frac{\partial^{#1}}{\partial {#2}^{#1}}} 
\newcommand{\set}[1]{\left\lbrace #1 \right\rbrace}
\newcommand{\iv}[1]{\operatorname{\text{d}#1}}
\newcommand{\mo}[1]{\hat{\operatorname{#1}}}

\makeatother

\begin{document}

\title{Non-equilibrium quantum transport coefficients and the transient
dynamics of full counting statistics in the strong coupling and non-Markovian
regimes}

\author{Javier Cerrillo, Maximilian Buser, Tobias Brandes}
\email{cerrillo@tu-berlin.de}

\affiliation{Institut f\"ur Theoretische Physik, Technische Universit\"at Berlin,
Hardenbergstr. 36, 10623 Berlin, Germany}
\begin{abstract}
Non-equilibrium transport properties of quantum systems have recently
become experimentally accessible in a number of platforms in so-called
full-counting experiments that measure transient and steady state
non-equilibrium transport dynamics. We show that the effect of the
measurement back-action can be exploited to establish general relationships between transport coefficients in the transient regime which take the form of fluctuation-dissipation theorems in the steady state. This result becomes most conspicuous in the transient dynamics of open quantum systems under strong coupling to non-Markovian environments in non-equilibrium settings. In order to explore
this regime, a new simulation method based in a hierarchy of equations
of motion has been developed. We
instantiate our proposal with the study of energetic conductance between
two baths connected via a few level system. 
\end{abstract}

\maketitle

\section{Introduction}
The experimental ability to probe the statistical properties of quantum
transport in mesoscopic systems, such as electrons in nanojunctions
\cite{Ubbelohde2012} or cold atoms \cite{Brantut2013}, has provided
new insights into the non-equilibrium behavior of quantum systems.
Theoretical tools for their description have been developed in the
form of so-called full counting statistics (FCS) methods \cite{Nazarov2003a,Esposito2009},
which constitute a generalization of the theory of moment and cumulant
generating functions and is conceptually based on a two-point measurement
scheme \cite{Esposito2009}, so that state projection of the bath after the first measurement is
automatically incorporated. The effects of this state collapse quickly fade in
Markovian baths, allowing for the derivation of steady-state fluctuation theorems that relate several transport coefficients governing the dynamics. In the linear response regime, these are none other than the celebrated fluctuation-dissipation theorems, such as the Kubo formula, or other properties such as the Onsager-Casimir relations. Far from equilibrium, these theorems can be generalized for non-linear coefficients \cite{Saito2008}.

Femtosecond laser pulses and other ultrafast control techniques provide access to the statistical response of quantum systems in the transient regime. The Jarzynski equality \cite{Jarzynski1997}, the Crooks theorem \cite{Crooks1999,Tasaki2000} and related relationships confirm that steady-state fluctuation-dissipation theorems need not automatically carry over to the transient regime. An interpretation of these effects as the consequence of measurement back-action in the bath supports the intuition that the failure of steady-state relations must be particularly evident in situations where the system-environment coupling is very large or the environmental evolution is so slow that non-Markovian effects become relevant. In this situation, only an explicit computation of the full dynamics is so far known to provide the correct insight into transport properties.

The behaviour of non-Markovian, strong coupling transport settings is captured
in the Levitov-Lesovik formula \cite{Levitov1993} in the case of
non-interacting particles. FCS for non-Markovian settings was studied
from a general perspective in \cite{Braggio2006,Flindt2008}, whereas
specific treatments include harmonic chains \cite{Agarwalla2011},
spin-boson or fermionic models in the perturbative regime \cite{Nicolin2011a,Carrega2014,Nietner2014,Gallego-Marcos2014,Schaller2014,Kaasbjerg2015,Carrega,Xue2015}
and general bosonic or fermionic systems for the first and/or second moments of the
dynamics \cite{Wang2015,Nicolin2011,Velizhanin2008,Velizhanin2010,Hutzen2012}. Bath statistics of open
quantum sysems has provided access to universal oscillations in high order cumulants
\cite{Flindt2009} and the Kondo signature in the spin-boson
model \cite{Saito2013} and fermionic models \cite{Smirnov2013}. FCS measurement
 strategies are attracting renewed attention \cite{Dasenbrook2016} together with optimized cumulant evaluation methods \cite{Benito2016}. Additionally, discussion of classical and quantum initial correlation
effects in shot noise has been addressed in \cite{Emary2011} and  thermodynamic consistency of FCS simulation methods has been studied in \cite{Hussein2014}.

Here we provide an alternative approach, which consists in quantitatively computing the deviation of the Saito-Utsumi coefficient relations \cite{Saito2008} when applied in transient situations. We show that this deviation is directly associated to a physical picture where no part of the system-environment compound is initially subject to measurement, and is expressed in terms of a natural symmetry that affects the cumulant generating function.
In order to investigate these effects, we have developed a simulation
method that incorporates FCS into the formalism of hierarchy of equations
of motion (HEOM) \cite{Tanimura1989}, an established method for the simulation
of general, multilevel open quantum systems (OQS) that is non-perturbative
in the coupling strength and which faithfully represents non-Markovian
effects of the environment. With this method we gain access to cumulants
of any desired order of environmental energetic and particle observables and arbitrary time-dependence of the Hamiltonian may be treated.
As a first example, we consider an open quantum system that couples
to a bosonic heat bath, although this procedure is equally valid for fermionic
baths. This method generalizes previous attempts that involved first
moments \cite{Jin2008,Kato2015} and to our knowledge it is the first
time that high order cumulants are simulated with this formalism.

We first introduce the FCS formalism and discuss a generalization
thereof that allows for the isolation of the transient measurement
back-action. A relationship between the back-action effect and transport
coefficients is presented that holds both close to equilibrium and
far from it. We further introduce the simulation method developed
for the investigation of these transient effects, which is based on
the HEOM formalism. We finally present simulation results on a bosonic
transport setting and show that the energetic conductance can be reliably
accessed through investigation of measurement back-action effects.

\section{Steady state coefficient relations} 
\begin{figure}
\includegraphics[width=1\columnwidth]{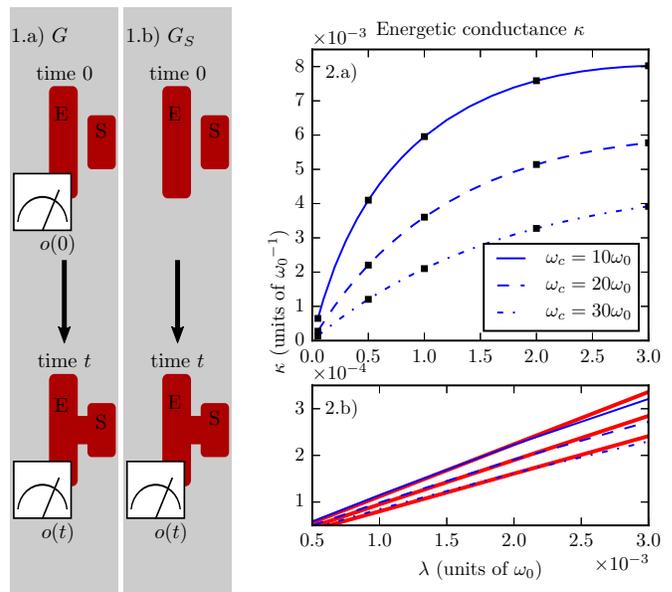}\protect
\caption{1) Depiction of measurement schemes associated to Eq.\eqref{GTM} (1.a) and Eq.\eqref{GS} (1.b) respectively.
In both measurement schemes a separable initial state of the form $\pi(A)$ is assumed.
2) Steady state energetic conductance as a function of the coupling strength
	$\lambda$ and for different spectral density cutoff frequencies $\omega_{c}$.
	(2.a) Lines show results computed via numerical derivation
	for $(T_{R}-T_{L})/T_{R}=0.01$. Dots are computed following Eq.\eqref{eq:CondNoBias}
	with no temperature bias. Other parameters are $T_{L}=T_{R}=10\omega_{0}$,
	$J=0$. All values have been numerically converged by varying the
	hierarchical depth of the simulation. (2.b) Closeup of the weak coupling
	limit, where red solid lines reproduce the analytical prediction of
	the weak coupling theory.}
\label{fig:Conductivity} 
\end{figure}

The formalism
for the analysis of full counting experiments constitutes a well-established
theoretical framework \cite{Esposito2009} involving a two point measurement
prescription: the value of a specific observable of interest $\mo O$
is measured at an initial time $t=0$, where a result $o\left(0\right)$
is obtained, and at a final time $t>0$, with an outcome $o\left(t\right)$ [see Fig.(\ref{fig:Conductivity}.1)].
Repetition of the experiment generates statistics of the measurement
difference $\Delta o\left(t\right)=o\left(t\right)-o\left(0\right)$,
which can be treated as a stochastic variable. Although more general cases can be considered, let us regard for simplicity a single measured operator and an initial state of the total system $\pi(A)=\rho\otimes\frac{\exp(-A\mo O)}{\tr{\exp(-A\mo O)}}$, where $\rho$ is an arbitrary state of the subsystem where the measurement has no effect and $A$ is a thermodynamic constraint fixing the initial expected value of $\mo O$. An instance of such a setting is a system in contact with several baths, where $\mo O$ is the Hamiltonian of one of the baths, $A$ is its initial inverse temperature and $\rho$ is an arbitrary state of the system and all non-measured baths.

Under these conditions, the cumulant generating function (CGF) takes the form \cite{Esposito2009}
\begin{equation}
G(\chi,A,t)=\ln\ev{e^{i\chi\mo O(t)}e^{-i\chi\mo O(0)}}_{A},\label{GTM}
\end{equation}
with $\ev{\bullet}_{A}\equiv\tr{\bullet\pi(A)}$. Its Taylor-expansion
coefficients in the counting field $\chi$
correspond to the cumulants of the measurement
difference $\Delta o\left(t\right)$. In the steady state limit, the cumulant generating function for the currents $\mathcal F (\chi,A)\equiv \lim_{\tau\rightarrow\infty} \frac{1}{\tau} G(\chi,A,\tau)$ often fulfills the symmetry
\begin{equation}
\mathcal F (\chi,A)=\mathcal F (-\chi+iA,A),
\label{FT}
\end{equation}
also known as fluctuation theorem, so that transport coefficients $L_m^n (A)\equiv\left.\frac{\partial^{m+n}}{\partial( i \chi)^n\partial A ^m}\mathcal F(\chi,A)\right|_{\chi=0}$ obey the Saito-Utsumi relations \cite{Saito2008}
\begin{equation}
L_m^n(A)=\sum_{j=0}^{m}\left(\begin{array}{c}
m\\
j
\end{array}\right) (-1)^{n+j} L_{m-j}^{n+j}(A).
\label{SU}
\end{equation}
Relations such as the Kubo formula and the Onsager-Casimir relations can be recast as specific cases of this equation, in particular when several counting fields $\chi_k$ are involved and the associated thermodynamic constraints are close to an equilibrium state $A_k\simeq A$. Nevertheless, Eq.(\ref{SU}) is generally not valid in the transient regime.

\section{Relations for transient transport coefficients} With the aim of quantifying the error of the fluctuation theorem Eq.\eqref{FT} in the transient dynamics, one may define the difference between two CGFs
\begin{equation}
G_{S}(\chi,A,t)\equiv\ln\ev{e^{i\chi\mo O(t)}}_{A}-\ln\ev{e^{i\chi\mo O(0)}}_{A},\label{GS}
\end{equation}
each associated to a single (S) measurement of the operator $\mo O$ given an initial state of the form $\pi(A)$ [see Fig.(\ref{fig:Conductivity}.1)]. Its $\chi-$derivatives provide the difference  of the cumulants of the measurement outcome between two times $t$ and $t=0$. Note the subtle difference in the statistical interpretation of functions Eq.\eqref{GTM} and Eq.\eqref{GS}.
As shown in Appendix \ref{sec:ApA}, both functions are related by the expression
\begin{equation}
G_S(\chi,A,t)=G(\chi,A-i\chi,t).
\end{equation}
Although this equation relates two physically distinct situations (two different measurement schemes), it bears a resemblance with Eq.(\ref{FT}) that can be exploited to obtain relations similar to Eq.(\ref{SU}) for the objects ${J_{(S)}}_m^n(A,t)\equiv\left.\frac{\partial^{m+n}}{\partial( i \chi)^n\partial A ^m}G_{(S)}(\chi,A,t)\right|_{\chi=0}$
\begin{equation}
{J_{S}}_m^n(A,t)=\sum_{j=0}^{n}\left(\begin{array}{c}
n\\
j
\end{array}\right) (-1)^{j} J_{m+j}^{n-j}(A,t).
\label{SUtran}
\end{equation}
Note that $L_m^n(A)=\lim_{\tau\rightarrow\infty} \frac{1}{\tau} J_m^n(A,\tau)$, and this can be used to recover Eq.(\ref{SU}) in the steady state.

This is a powerful
relationship that establishes analogies of fluctuation-dissipation theorems on the transient dynamics by quantifying the deviation from the steady state expressions in terms of the values ${J_{S}}_m^n(A,t)$, which correspond to measurable quantities in a well defined physical setting. The amplitude of these deviations grows with the strength of the coupling between the measured partition and the rest of the system, and also with the duration of the transient dynamics, which is associated to non-Markovian effects in the language of open quantum systems. We explore below these effects at hand of specific examples.

\begin{figure}
\includegraphics[width=1\columnwidth]{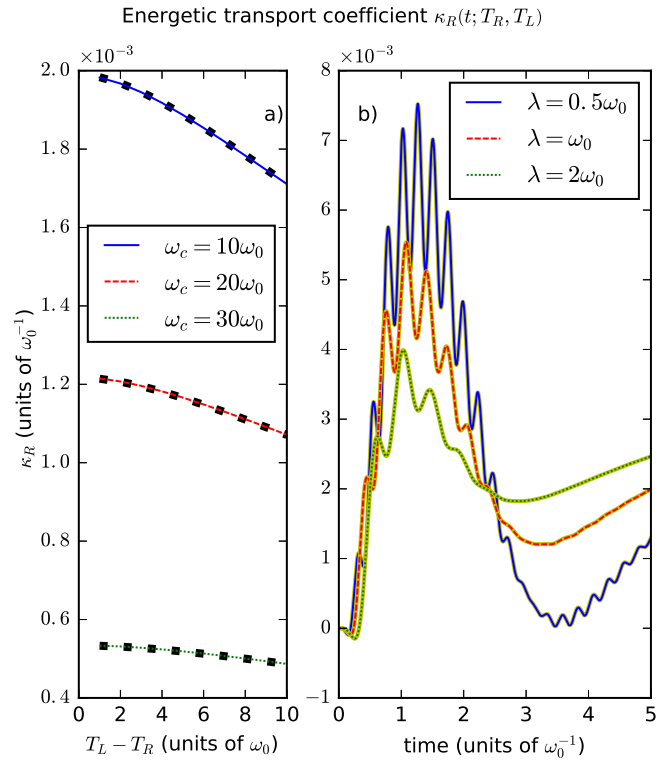}\protect\caption{(a) Steady state energetic transport coefficient $\lim_{t\rightarrow\infty}\kappa_{R}\left(t;T_{R},T_{L}\right)$
as a function of the temperature of the left bath $T_{L}$ and for
different spectral density cutoff frequencies $\omega_{c}$. Other
parameters are $T_{R}=10\omega_{0}$, $J=\omega_{0}$, $\lambda=\omega_{0}$.
Lines show results computed via
numerical derivation of steady-state currents. Dots are computed following
Eq.\eqref{eq:CondNoBias}. (b) Transient dynamics of the energetic
transport coefficient as a function of time and for different reorganization
energies $\lambda$. Other parameters are $T_{R}=T_{L}=10\omega_{0}$,
$J=\omega_{0}$, $\omega_{c}=3\omega_{0}$. Results shown in blue
are computed via numerical derivation
of transient currents. Red lines are computed following Eq.\eqref{eq:CondNoBias}.}
\label{fig:Transient} 
\end{figure}

\section{Applications} In order to instantiate the relations in Eq.(\ref{SUtran}), let us consider an open quantum system of Hamiltonian $\mo{H}_{\mathrm{S}}$ interacting via operators $\mo{V}_{\nu}$ with
one or several baths, so that the total Hamiltonian has the form
\begin{align}
\mo{H} & =\mo{H}_{\mathrm{S}}+\sum_{\nu}\mo{V}_{\nu}\otimes\mo{B}_{\nu}+\mo{H}_{\mathrm{B}},\label{eq:Ham}\\
\mo{H}_{\mathrm{B}} & =\sum_{\nu}\mo{H}_{\nu}=\sum_{\nu,k}\omega_{\nu,k}\mo{a}_{\nu k}^{\dagger}\mo{a}_{\nu k}.
\end{align}
where $\mo{B}_{\nu}$ is an arbitrary operator of bath $\nu$ and $\mo{a}_{\nu,k}$ and $\mo{a}_{\nu,k}^{\dagger}$
are the usual bosonic or fermionic annihilation and creation operators
for the mode $k$ of frequency $\omega_{\nu,k}$.
This encompasses a broad range of dissipative and transport settings. 

The energy of one of the baths $\mo{H}_{\nu}$
provides relevant information about the heat flows in the system.
The thermodynamic constraint $\beta_{\nu}$ associated to it is the inverse temperature of the bath, so that Eq.\eqref{SUtran} for $m=0$ takes the form
\begin{equation}
{J_{S}}_0^n(\beta_\nu,t)=\sum_{j=0}^{n}\left(\begin{array}{c}
n\\
j
\end{array}\right) (-1)^{j} J_{n-j}^{j}(\beta_\nu,t),\label{eq:FCSrel-1}
\end{equation}
which relates the cumulants of the bath energies with high order instantaneous
energetic conductances. It has the same form as the case $n=0$ of Eq.(\ref{SU}) except for the {\em deviation} term ${J_{S}}_0^n$ and an extra minus sign for odd $n$.

For a two-bath setting ($\nu\in\left\{ R,L\right\} $),
we define a first order \emph{energetic transport coefficient} as $\kappa_{R}\left(t;T_{R},T_{L}\right)\equiv \beta^{2}_R \prt{}{t} J_1^1(\beta_R,t),$\footnote{Please note the special choice of sign, such that subsequent numerical results remain positive.}
whose steady state, equilibrium limit \emph{energetic conductance}
$\kappa\equiv\lim_{t\rightarrow\infty}\lim_{T_R\rightarrow T_L}\kappa_{R}\left(t;T_R,T_L\right)$ can be understood in turn as a generalization of the concept of
thermal conductance for the case of strong-coupling and non-Markovian
regimes. Equation \eqref{eq:FCSrel-1} for $n=2$ establishes a relationship
between the first order energetic transport coefficient and the second order cumulants of the form 
\begin{equation}
\kappa_{R}\left(t;T_{R},T_{L}\right)=\frac{\beta^2_R}{2} \prt{}{t}\left[J_0^2\left(\beta_R,t\right)-{J_{S}}_0^2\left(\beta_R,t\right)\right],\label{eq:CondNoBias}
\end{equation}
which is analogous to the Kubo formula except for the deviation term ${J_{S}}_0^2$. As in the case of the fluctuation-dissipation theorem, it puts forward the possibility to derive a conductance value from the difference of the fluctuations as obtained from a two measurement scheme and a single measurement scheme. This is a way to circumvent the necessity of a numerical derivative and avoids
the accumulation of numerical error implicit in the choice of any
small but finite temperature bias.

\section{Numerical method} In order to address the strong-coupling and non-Markovian
regimes, we developed a hierarchy of equations of motion for the simulation of full
counting statistics and high-order moments of relevant bath observables. The technique
enables statistical analysis of any bath observable that commutes with its
free Hamiltonian term, and we
provide examples for the case of the energy of the
bath $\mo{H}_{\mathrm{B}}$. In particular, any linear combination of the operators $\mo{a}_{\nu k}^{\dagger}\mo{a}_{\nu k}$
such as the particle number $\mo{N}$ can be simulated with this approach. Furthermore, it is flexible enough to generate
both the moments corresponding to the two-measurement picture Eq.\eqref{GTM}
and the single measurement picture Eq.\eqref{GS}, and we will use it in
both modes and apply the relationship Eq.\eqref{SUtran} to obtain
non-equilibrium transport coefficients.

A central element in the derivation of the method is the counting-field-resolved bath correlation function
\begin{equation}
C_{\nu}^{jk}(\chi,t)=(-)^{j+k}\ev{\tilde{\mo{B}}_{\nu}^{j}\left[(-)^{j}\frac{\chi}{2}\right](t)\tilde{\mo{B}}_{\nu}^{k}\left[(-)^{k}\frac{\chi}{2}\right](0)},\label{eq:CFchi-1}
\end{equation}
where superindices $j$ and $k$ take two values $0$ or $1$ and indicate the side an operator acts from: $\mo{A}^{0}\rho\equiv\mo{A}\rho$ or $\mo{A}^{1}\rho\equiv\rho\mo{A}$. The tilde indicates the interaction picture with respect to $\mo{H}_{\nu}$, the dressing of an operator by the counting operator $\mo{O}$ is denoted by $\mo{A}[\chi](t)=e^{i\chi\mo{O}}\mo{A}(t)e^{-i\chi\mo{O}}$ and the average is over the initial state of the bath.
The hierarchy is based on a decomposition of the correlation function $C_{\nu}^{jk}(\chi,t)=\sum_{r}c_{\nu r}^{jk}(\chi)\phi_{r}(t)$
by means of a set of functions $\{\phi_{r}(t)\}$ whose derivatives
are well defined within the set by $\de{}{t}\phi_{r}(t)=\sum_{r}\eta_{rs}\phi_{s}(t)$ \cite{Tang2015}. The set size determines the growth of the simulation requirements, traditionally making this method indicated for not too low temperature regimes. Nevertheless, a judicious choice of the basis can circumvent this limitation \cite{Tang2015}.

\begin{figure}
\includegraphics[width=1\columnwidth]{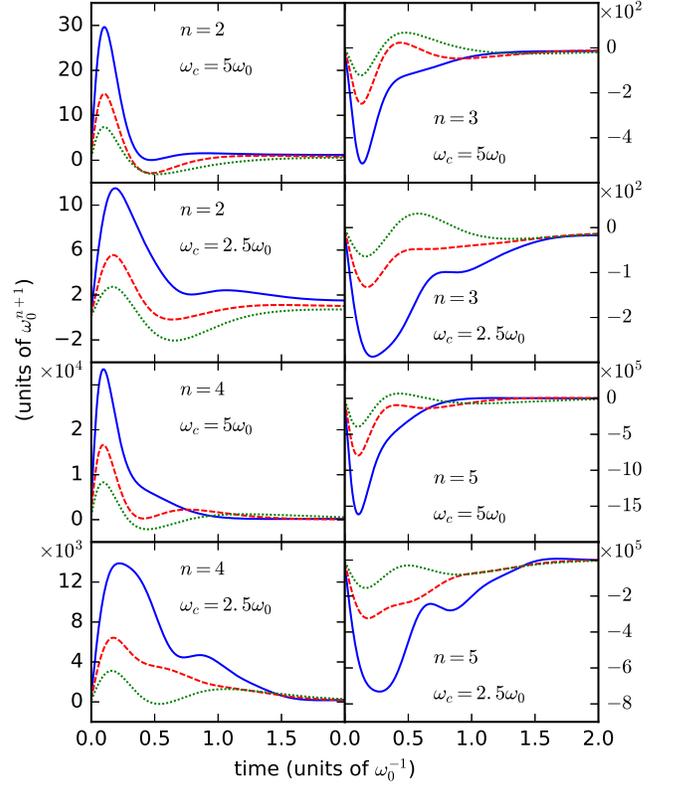}\protect\caption{Transient dynamics of the time derivatives of the first 5 cumulants
of the bath energy $\mo{H}_{R}$ in the single measurement scheme (S)
for different values of  the cutoff frequency $\omega_{c}$ and the reorganization energy $\lambda$: $\lambda=0.05\omega_c$ for the dotted lines, $\lambda=0.1\omega_c$ for the dashed lines and $\lambda=0.2\omega_c$ for the solid lines. Other parameter values are: $T_{L}=T_{R}=10\omega_{0}$,
$J=0$.}
\label{fig:5Cum} 
\end{figure}

We refer the reader to the Appendices (sections B \& C) for the technical aspects of the method. It involves the propagation of so-called {\em auxiliary fields}
$\sigma_{\set{m}}^{\set{n}}\left(t\right)$, labeled by a
rank-three tensor  $\set{n}$ and a vector $\set{m}$ of non-negative integer entries whose elements respectively sum up to the hierarchic level $n$ and represent a partition of the moment order $m$. 
The zeroth hierarchic level contains the auxiliary field $\sigma_{\set{0}}^{\set{0}}\left(t\right)$,
corresponding to the system density matrix $\rho\left(t\right)$, and $\sigma_{\set{m}}^{\set{0}}\left(t\right)$, a linear combination of which (defined in the Appendix, Eq.\ref{mom}) can be traced to obtain the $m$th moment.
The auxiliary fields satisfy the equation 
\begin{align}
\de{}{t}\sigma_{\set{m}}^{\set{n}} & (t)=-i\mo{H}_{S}^{\times}\sigma_{\set{m}}^{\set{n}}(t)\nonumber \\
+ & \sum_{\nu,r;k=0,1}\Bigg(\mo{\bar{V}}_{\nu r}^{k}\sigma_{\set{m}}^{\set{\dots,n_{\nu r}^{k}+1,\dots}}(t)\nonumber \\
+ & \sum_{s}n_{\nu r}^{k}\eta_{rs}\sigma_{\set{m}}^{\set{\dots,n_{\nu r}^{k}-1,\dots,n_{\nu s}^{k}+1,\dots}}(t)\nonumber \\
+ & n_{\nu r}^{k}\phi_{r}(0)\mo{V}_{\nu}^{k}\sigma_{\set{m}}^{\set{\dots,n_{\nu r}^{k}-1,\dots}}(t)\nonumber \\
+ & \sum_{q}m_{q}\mo{\bar{V}}_{\nu rq}^{k}\sigma_{\set{\dots,m_{q}-1,\dots}}^{\set{\dots,n_{\nu r}^{k}+1,\dots}}(t)\Bigg),\label{HEOMOBS-1}
\end{align}
where $\mo{\bar{V}}_{\nu r}^{k}\equiv\sum_{j=0,1}c_{\nu r}^{jk}(0)\mo{V}_{\nu}^{j}$, $\mo{\bar{V}}_{\nu rq}^{k}\equiv\sum_{j=0,1}c_{\nu rq}^{jk}\mo{V}_{\nu}^{j}$
and $c_{\nu rq}^{jk}\equiv\left.\de{q}{(i\chi)}c_{\nu r}^{jk}(\chi)\right|_{\chi=0}$.
The structure is identical to the usual hierarchy \cite{Tanimura1989}
but for the last term, which connects it to the hierarchy associated to the previous moment, which is compatible with results in the Markovian, weak-coupling case \cite{Benito2016}. A cutoff at a maximum hierarchic level $n_{max}$ is usually justified in terms of numerical convergence and is roughly proportional to the system-bath coupling strength.

In order to demonstrate the numerical validity of Eq.\eqref{eq:CondNoBias},
let us consider a two level system with Hamiltonian $\mo{H}_{\mathrm{S}}=\omega_{0}\sigma_{x}+J\sigma_{z}$, where $\sigma_{i}$ with $i\in\left\{ x,y,z\right\} $ are the Pauli matrices, coupled to two bosonic baths $\nu\in\{R,L\}$ via $\mo{V}_{\nu}=\sigma_z$ and $
\mo{B}_{\nu}=\sum_{k}\gamma_{\nu k}\left(\mo{a}_{\nu k}+\mo{a}_{\nu k}^{\dagger}\right)$, characterized by a spectral density $J_{\nu}(\omega)=\sum_{k}\gamma^{2}_{\nu k}\delta(\omega-\omega_{\nu k})$
and an inverse temperature $\beta_{\nu}$. The following results are
derived for $\rho(0)=\frac{\sigma_x+1}{2}$ and the choice of an Ohmic spectral density with exponential
cutoff $J(\omega)=\frac{\lambda}{\omega_{c}}\omega e^{-\frac{\omega}{\omega_{c}}}$,
where $\lambda=\int_{0}^{\infty}\frac{J\left(\omega\right)}{\omega}d\omega$
is the reorganization energy and $\omega_{c}$ is the scale of the
cutoff.

Energetic conductance values are shown in Fig.(\ref{fig:Conductivity}.b).
Following the discussion presented above, two methods are used for
their computation. On the one hand, simulations under small temperature
bias $(T_{R}-T_{L})/T_{R}=0.01$ are run in order to obtain steady
state energy currents, which are related to the energetic conductance
through the definition of $J_1^1(\beta_R,t)$. On the other hand,
second order moments computed in the two-measurement  and the single measurement schemes for equilibrium conditions
 ($T_{L}=T_{R}$) are further used to derive conductance
values as per Eq.\eqref{eq:CondNoBias}. Results show excellent agreement
between both pictures and also approach predictions from
the weak coupling theory and the {\em non-interacting blip approximation} (NIBA) \cite{Nicolin2011a} in their respective regimes of
validity. Whereas a linear increase of conductance with the coupling
strength $\lambda$ is expected in the weak coupling limit, a turnover
is reproduced for higher coupling strengths. The relationship Eq.\eqref{eq:CondNoBias}
is valid also in the transient regime and far from equilibrium as
shown in Fig.\eqref{fig:Transient}. In this case, the steady state
energetic conductance of a biased two level system ($J\neq0$) is
studied in situations where $T_{R}-T_{L}\simeq T_{L}$. The transient
dynamics are also accurately reproduced by Eq.\eqref{eq:CondNoBias},
where the oscillating effect is introduced by the tunneling, which
acts as an effective driving.

Finally, the transient deviation of high order transport coefficient relations Eq.(\ref{SU}) is shown in Fig.(\ref{fig:5Cum}) for a range of coupling
constants $\lambda$ and spectral density cutoffs $\omega_{c}$. It constitutes a quantitative confirmation that the failure of steady-state fluctuation-dissipation theorems is proportional to the coupling strength to the bath and lasts longer for higher degrees of non-Markovianity. Additionally, the sign of the deviation changes for odd orders, as predicted from the comparison of Eq.(\ref{SU}) and Eq.\eqref{eq:CondNoBias}.

\section{Conclusions} When considering the transient
transport dynamics of non-Markovian systems, it is possible to quantify deviations from fluctuation-dissipation theorems and the nonlinear Saito-Utsumi relations, which are only valid in the steady state. These deviations have a physical interpretation and are associated to equilibration dynamics of the same system under a different measurement scheme. We demonstrate this relation by developing a tool
that allows for the simulation of the full counting statistics of
a broad range of bath observables under dissipative and non-equilibrium
settings, which is a generalization of the celebrated hierarchy of
equations of motion for non-Markovian and strong-coupling settings.
By accessing high order cumulants of the bath energy, it is possible
to derive energetic conductances and higher order derivatives thereof
while, at the same time, avoiding finite bias simulations.
This approach is immediately applicable to the study of observables
such as the particle number, environments of fermionic
nature, more complex
and higher dimensional systems, and time-dependent driving.

\acknowledgements We are grateful to Dr. Gernot Schaller
and Dr. Philipp Strasberg for helpful discussions. The authors gratefully
acknowledge financial support from the DFG Grants No. BR 1528/7, No.
BR 1528/8, No. BR 1528/9, No. SFB 910 and No. GRK 1558.

\appendix

\section{Relevant limits of the cumulant generating function\label{sec:ApA}}

Let us reproduce the definition of the cumulant generating function
\begin{align}
G(\chi,A,t)&\equiv\ln\ev{e^{i\chi\mo O}(t)e^{-i\chi\mo O}(0)}_{A}\nonumber\\
&=\ln \tr{e^{i\chi\mo O}(t)e^{-i\chi\mo O}(0)\pi\left(A\right)},\label{GTMApp}
\end{align}
with $\pi\left(A\right)=\rho\otimes \frac{e^{-A\mo O}}{\tr{e^{-A\mo O}}}$ a separable initial state of the total system, where $\rho$ is an arbitrary state of the subsystem where the measurement has no effect and $A$ is a thermodynamic constraint fixing the initial expected value of $\mo O$. Evaluating the second variable of the cumulant generating function at $A-i\chi$ one obtains
\begin{align}
G(\chi,&A-i\chi,t)=\nonumber\\
&=\ln \tr{e^{i\chi\mo O}(t)\pi\left(A\right)}-\ln \frac{\tr{e^{(-A+i\chi)\mo O}}}{\tr{e^{-A\mo O}}}\nonumber\\
&=\ln\ev{e^{i\chi\mo O}(t)}_{A}-\ln\ev{e^{i\chi\mo O}(0)}_{A}.
\end{align}
One may interpret this function as the difference between two single-measurement cumulant generating functions: one where the measurement takes place at time $t$ and another one where the measurement takes place at time $t=0$. We denote it by
\begin{equation}
G_S(\chi,A,t)\equiv\ln\ev{e^{i\chi\mo O}(t)}_{A}-\ln\ev{e^{i\chi\mo O}(0)}_{A}.\label{QCGF}
\end{equation}

\section{Counting-field-resolved hierarchy of equations of motion\label{FOM}}

As indicated in the main text Eq.(7-8), let us consider a general total
Hamiltonian consisting of a system, interaction and bath parts 
\begin{align}
\mo{H} & =\mo{H}_{\mathrm{S}}+\sum_{\nu}\mo{V}_{\nu}\otimes\mo{B}_{\nu}+\mo{H}_{\mathrm{B}},\\
\mo{H}_{\mathrm{B}} & =\sum_{\nu}\mo{H}_{\nu}=\sum_{\nu,k}\omega_{\nu,k}\mo{a}_{\nu k}^{\dagger}\mo{a}_{\nu k}.
\end{align}
The index $\nu$ labels the baths, $\mo{a}_{\nu,k}$ and $\mo{a}_{\nu,k}^{\dagger}$
are the usual bosonic or fermionic annihilation and creation operators
for the mode $k$ in bath $\nu$. The goal is to derive the full counting
cumulant generating function Eq.\eqref{GTM} of an observable $\mo{O}$
of one of the baths which commutes with its free Hamiltonian $\mo H_{\nu}$
and with the initial state of the system and baths $\pi\left(0\right)=\rho(0)\bigotimes_{\nu}{\frac{e^{-\beta_\nu\mo H_\nu}}{\tr{e^{-\beta_\nu \mo H_\nu}}}}$. For simplicity, we will omit the explicit dependence on the thermodynamic constraint in the derivation, so that the two-measurement cumulant generating function takes the form
\begin{equation}
G(\chi,t)=\ln\tr{e^{-i\mo{H}\left[\frac{\chi}{2}\right]t}\pi(0)e^{i\mo{H}\left[-\frac{\chi}{2}\right]t}},\label{G2-2}
\end{equation}
where $\mo{A}[\chi](t)=e^{i\chi\mo{O}}\mo{A}(t)e^{-i\chi\mo{O}}$.
This problem can be formulated in terms of the solution to a hierarchy
of equations of motion for the counting field resolved density matrix
\begin{equation}
\rho(\chi,t)=\tra{B}{e^{-i\mo{H}\left[\frac{\chi}{2}\right]t}\pi(0)e^{i\mo{H}\left[-\frac{\chi}{2}\right]t}},\label{eq:DMchi}
\end{equation}
and the equation $G(\chi,t)=\ln\left[\tr{\rho(\chi,t)}\right]$
directly relates both quantities.

For the sake of clarity, we will derive the equations of motion for
Eq.\eqref{eq:DMchi}.
The case associated to Eq.\eqref{QCGF} immediately follows under modification of
the initial state of the bath with the imaginary inverse temperature $\beta_\nu -i\chi$.
The matrix $\rho(\chi,t)$ satisfies the differential equation 
\begin{equation}
\frac{d}{dt}\rho(\chi,t)=-i\tra{B}{\mo{H}\left[\frac{\chi}{2}\right]\pi(\chi,t)-\pi(\chi,t)\mo{H}\left[-\frac{\chi}{2}\right]},\label{vN}
\end{equation}
with $\pi(\chi,t)\equiv e^{-i\mo{H}\left[\frac{\chi}{2}\right]t}\pi(0)e^{i\mo{H}\left[-\frac{\chi}{2}\right]t}$.
The formal solution of Eq.(\ref{vN})
in the interaction picture with respect to $\mo{H}_{\mathrm{S}}+\mo{H}_{\mathrm{B}}$
(denoted by an overhead tilde) can be obtained by means of Wick's theorem. For simplicity, we will focus on the derivation for the bosonic case, but all steps can be trivially generalized for the fermionic case. Wick's theorem  simplifies the calculation of the partial trace of the bath by reducing products of 2n  operators to n  products of pairwise traces
 \begin{align}
\left\langle \hat{\mathrm{T}}\tilde{\hat{\mathrm{B}}}_{\nu}\left(t_{2n}\right)\tilde{\hat{\mathrm{B}}}_{\nu}\left(t_{2n-1}\right)\cdots\tilde{\hat{\mathrm{B}}}_{\nu}\left(t_{2}\right)\tilde{\hat{\mathrm{B}}}_{\nu}\left(t_{1}\right)\right\rangle =\nonumber\\
\sum_{app}\prod_{ij}\left\langle \hat{\mathrm{T}}\tilde{\hat{\mathrm{B}}}_{\nu}\left(t_{i}\right)\tilde{\hat{\mathrm{B}}}_{\nu}\left(t_{j}\right)\right\rangle ,
\end{align}
 where the sum is over all possible pairs (app) of indices up to $2n$, $\mo{T}$
is the time ordering operator and the average may be performed over any Gaussian state. Therefore, the solution of Eq.\eqref{vN} in the interaction picture is $\tilde{\rho}(\chi,t)=\tilde{\mathcal{U}}(\chi,t)\rho(0,0)$,
with 
\begin{equation}
\tilde{\mathcal{U}}(\chi,t)=\prod_{\nu}\prod_{jk=0,1}\exp_{+}\left(\int\limits _{0}^{t}\iv{s}\tilde{\mo{W}}_{\nu}^{jk}(\chi,s)\right),
\end{equation}
where $\exp_{+}$ stands for the time ordered exponential and 
\begin{equation}
\tilde{\mo{W}}_{\nu}^{jk}(\chi,t)=-\int\limits _{0}^{t}\iv{s}\tilde{\mo{V}}_{\nu}^{j}(t)C_{\nu}^{jk}(\chi,t-s)\tilde{\mo{V}}_{\nu}^{k}(s).
\end{equation}
Here we introduce the superoperator notation $\mo{A}^{0}\rho\equiv\mo{A}\rho$
and $\mo{A}^{1}\rho\equiv\rho\mo{A}$ and the bath correlation functions
\begin{equation}
C_{\nu}^{jk}(\chi,t)=(-)^{j+k}\ev{\tilde{\mo{B}}_{\nu}^{j}\left[(-)^{j}\frac{\chi}{2}\right](t)\tilde{\mo{B}}_{\nu}^{k}\left[(-)^{k}\frac{\chi}{2}\right](0)},\label{eq:CFchi}
\end{equation}
where $\ev{A}=\tr{A\pi(0)}$. Note that this definition should be
replaced by $\ev{A}=\tr{A\pi(0)e^{i\chi\mo{O}}\tr{e^{i\chi\mo{O}}\pi(0)}^{-1}}$
for Eq.\eqref{QCGF}.

The HEOM formalism uses an approximate representation of the correlation
functions by means of linear combinations of decaying exponential
functions. The extended version generalizes the framework to more
involved functional bases \cite{Tang2015}. In our case, the coefficients
of the linear combination are functions dependent on the counting
field $\chi$, so that we approximate $C_{\nu}^{jk}(\chi,t)=\sum_{r}c_{\nu r}^{jk}(\chi)\phi_{r}(t)$
by means of a set of functions $\{\phi_{r}(t)\}$ whose derivatives
are well defined within the set by $\de{}{t}\phi_{r}(t)=\sum_{n}\eta_{rs}\phi_{s}(t)$, where $\eta$ is a matrix with complex entries. The form of $c_{\nu r}^{jk}\left(\chi\right)$
  is general and depends on the specific observable of interest $\hat{\mathrm{O}}$. For instance, in the case $\hat{\mathrm{O}}=\mathrm{H}_{\nu}$ and $
\mo{B}_{\nu}=\sum_{k}\gamma_{\nu k}\left(\mo{a}_{\nu k}+\mo{a}_{\nu k}^{\dagger}\right)$, $C_{\nu}^{jk}\left(\chi,t\right)=C_{\nu}^{jk}\left(\chi\pm t\right)$
  and the dependence is expected to be similar to that of $\phi_{r}\left(t\right)$.
With this representation, it is possible to define the auxiliary objects
\begin{align}
&\tilde{\rho}^{\set{n}}(\chi,t)=\nonumber\\
&\mo{\text{T}}\prod\limits _{\nu,r;k=0,1}\left(\int\limits _{0}^{t}\iv{s}\phi_{r}\left(t-s\right)\tilde{\mo{V}}_{\nu}^{k}(s)\right)^{n_{\nu r}^{k}}\tilde{\mathcal{U}}(\chi,t)\rho(0,0),
\end{align}
where $\set{n}\equiv\set{\dots,n_{\nu r}^{k},\dots}$ is a rank-three tensor of non-negative integer entries which sum up to $n$ and $n$ is the so-called {\em hierarchic level}. It is clear that $\tilde{\rho}^{\set{0}}(\chi,t)=\tilde{\rho}(\chi,t)$
and the auxiliary fields satisfy the equation 
\begin{alignat}{1}
\de{}{t}\rho^{\set{n}}(\chi,t)= & -i\mo{H}_{S}^{\times}\rho^{\set{n}}(\chi,t)\\&+\sum_{\nu,r;k=0,1}\Bigg(\bar{\mo{V}}_{\nu r}^{k}\rho^{\set{\dots,n_{\nu r}^{k}+1,\dots}}(\chi,t)\nonumber\\&+\sum_{s}n_{\nu r}^{k}\eta_{rs}\rho^{\set{\dots,n_{\nu r}^{k}-1,\dots,n_{\nu s}^{k}+1,\dots}}(\chi,t)\label{HEOMFCS}\nonumber\\
 & +n_{\nu r}^{k}\phi_{r}(0)\mo{V}_{\nu}^{k}\rho^{\set{\dots,n_{\nu r}^{k}-1,\dots}}(\chi,t)\Bigg),\nonumber 
\end{alignat}
where we have used the notation $\mo{A}^{\times}\rho\equiv\mo{A}\rho-\rho\mo{A}$
and $\bar{\mo{V}}_{\nu r}^{k}\equiv\sum_{j=0,1}c_{\nu r}^{jk}(\chi)\mo{V}_{\nu}^{j}$.
This is an extension of the usual HEOM formulation \cite{Tanimura1989}.

\section{Hierarchy of equations of motion for high order statistical moments\label{sub:HH}}

Although Eq.\eqref{HEOMFCS} can be used on its own to obtain the
generating function, in the case where one is interested in specific
statistical moments, a specialized hierarchy can be derived. Let us
define the object 
\begin{align}
\tilde{\sigma}_{m}(t)&\equiv \left.\prt{m}{(i\chi)}\tilde{\rho}^{\set{0}}(\chi,t) \right|_{\chi=0}\nonumber\\&=\mo{\text{T}}\left.\prt{m}{(i\chi)}\tilde{\mathcal{U}}(\chi,t)\right|_{\chi=0}\rho(0,0),
\end{align}
so that the moment $m$ of the full counting distribution may be obtained by tracing: $\tr{\tilde{\sigma}_{m}(t)}$.
It contains correlation
functions of the form $\left.\prt{q}{(i\chi)}C_{\nu}^{jk}(\chi,t)\right|_{\chi=0}\equiv C_{\nu q}^{jk}(t)$,
which are well defined in terms of the approximate representation
as $C_{\nu q}^{jk}(t)=\sum_{r}c_{\nu rq}^{jk}\phi_{r}(t)$ with $c_{\nu rq}^{jk}\equiv\left.\de{q}{(i\chi)}c_{\nu r}^{jk}(\chi)\right|_{\chi=0}$.
In a procedure analogous to the one followed for the obtention of
Eq.(\ref{HEOMFCS}), we define 
\begin{align}
\tilde{\sigma}_{\set{m}}^{\set{n}}(t)&=\mo{\text{T}}\prod\limits _{\nu,k;j=0,1}\left(\int\limits _{0}^{t}\iv{s}\phi_{k}\left(t-s\right)\tilde{\mo{V}}_{\nu}^{j}(s)\right)^{n_{\nu k}^{j}}\nonumber\\&\prod_{q}\left(\int\limits _{0}^{t}\iv{s}\tilde{\mo{W}}_{\nu q}(s)\right)^{m_{q}}\tilde{\mathcal{U}}(0,t)\rho(0,0),
\end{align}
where $\set{m}\equiv\set{\dots,m_{q},\dots}$ is a vector of non-negative integer entries such that $\sum_{q}m_{q}q=m$ and $\tilde{\mo{W}}_{\nu q}(t)\equiv\sum_{j,k=0,1}-\int\limits _{0}^{t}\iv{s}\tilde{\mo{V}}_{\nu}^{j}(t)C_{\nu q}^{jk}(t-s)\tilde{\mo{V}}_{\nu}^{k}(s)$.
This object satisfies the equation 
\begin{align*}
\de{}{t}\sigma_{\set{m}}^{\set{n}}(t)= & -i\mo{H}_{S}^{\times}\sigma_{\set{m}}^{\set{n}}(t)\\&+\sum_{\nu,r;k=0,1}\Bigg(\bar{\mo{V}}_{\nu r}^{k}\sigma_{\set{m}}^{\set{\dots,n_{\nu r}^{k}+1,\dots}}(t)\\&+\sum_{s}n_{\nu r}^{k}\eta_{rs}\sigma_{\set{m}}^{\set{\dots,n_{\nu r}^{k}-1,\dots,n_{\nu s}^{k}+1,\dots}}(t)\\
 & +n_{\nu r}^{k}\phi_{r}(0)\mo{V}_{\nu}^{k}\sigma_{\set{m}}^{\set{\dots,n_{\nu r}^{k}-1,\dots}}(t)\\&+\sum_{q}m_{q}\bar{\mo{V}}_{\nu rq}^{k}\sigma_{\set{\dots,m_{q}-1,\dots}}^{\set{\dots,n_{\nu r}^{k}+1,\dots}}(t)\Bigg),
\end{align*}
where $\bar{\mo{V}}_{\nu rq}^{k}\equiv\sum_{j=0,1}c_{\nu rq}^{jk}\mo{V}_{\nu}^{j}$.
The structure is identical to the usual hierarchy (Eq. \ref{HEOMFCS})
but for the last term, which connects it to the next tier elements
of the hierarchy associated to the previous moment. This can be interpreted
as an additional driving that each moment exerts on the next one.
This naturally defines a cascade of hierarchies that can be exploited
for parallel simulation of several moments with reduced overhead.

The relationship between $\tilde{\sigma}_{m}(t)$ and the set of $\tilde{\sigma}_{\set{m}}^{\set{n}}(t)$
follows 
\begin{equation}
\tilde{\sigma}_{m}(t)=\sum_{\set{m}}a_{\set{m}}\tilde{\sigma}_{\set{m}}^{\set{0}}(t).
\label{mom}
\end{equation}
where the sum is over all partitions $\set{m}$ of $m$ (all vectors $m_{q}$
with the property $\sum_{q=1}^{m}m_{q}q=m$) and 
\[
a_{\set{m}}\equiv\prod_{q=1}^{m}\prod_{j=1}^{m_{q}}\frac{1}{j}\left(\begin{array}{c}
m-\sum_{r=q}^{m}rm_{r}+jq\\
k
\end{array}\right)
\]
 is the number of permutations associated to that partition.

Finally, cumulants recursively relate to moments by means of the formula
\begin{equation}
\ev{\ev{A^{n}}}=\ev{A^{n}}-\sum_{m=1}^{n-1}\left(\begin{array}{c}
n-1\\
m-1
\end{array}\right)\ev{\ev{A^{m}}}\ev{A^{n-m}}.
\end{equation}
Furthermore factorial cumulants are obtained by
\begin{equation}
\ev{\ev{A^{n}}}_{F}=\ev{\ev{A^{n}}}-\sum_{m=1}^{n-1}\left\{ \begin{array}{c}
n\\
m
\end{array}\right\} \ev{\ev{A^{m}}}_{F},
\end{equation}
where $\left\{ \begin{array}{c}
n\\
m
\end{array}\right\} $ are the Stirling numbers of the second kind.

\bibliographystyle{apsrev4-1}

\end{document}